# The effect of the curvature-scalar mixing on the annihilation cross sections of dark matter scalar and unparticle dark matter

E. O. Iltan [*]

Physics Department, Middle East Technical University
Ankara, Turkey

**Abstract**

We consider the dark matter scalar in the Randall Sundrum background and study the annihilation cross section if the curvature-scalar mixing is switched on. In this case, in addition to the radion, the standard model Higgs scalar drives the annihilation and it leads to a considerable enhancement in the annihilation cross section. Furthermore we take unparticle, having a non-zero mass coming from the standard model Higgs-unparticle interaction in the low energy level, as a dark matter candidate and analyze the annihilation cross section by including the effect of the curvature-scalar mixing. We see that, for both choices, the mixing process plays an essential role in obtaining the dark matter annihilation cross section in the current range.

---

[*]E-mail address: eiltan@newton.physics.metu.edu.tr

The search for a theoretical background to explain the invisible matter, the so called dark matter (DM), reaches great interest since the DM contributes almost 23% of present Universe [1]-[3] with numerous evidences, the galactic rotation curves [4], galaxies orbital velocities [5], the cosmic microwave background anisotrophy [6], the observations of type Ia supernova [3]. The DM problem can not be solved in the framework of the standard model (SM) and one needs to go beyond. The possible scenarios are the Supersymmetry [7], the universal and non universal extra dimension (UED and NUED) models [8]-[20], the split UED models [21]-[23], the Private Higgs model [24], the Inert doublet model [25]-[31], the Little Higgs model [32], the Heavy Higgs model [33].

The common belief is that the Weakly Interacting Massive Particles (WIMPs), made of cold relics, are DM candidates. These particles have masses in the range 10 GeV- a few TeV and are required to be stable, having no decay products and playing a crucial role in the structure formation of the Universe. From the theoretical point of view, the stability is ensured by a discrete symmetry which is model dependent. In the supersymmetric models the R parity, in the models with extra dimensions the Kaluza-Klein (KK) parity, in the Little Higgs models the T-parity are the discrete symmetries ensuring the stability of the DM candidate. If the DM is taken as an additional scalar in the model under consideration, its stability is guaranteed by introducing an ad-hoc $Z_2$ symmetry in the lagrangian and it disappears by the annihilation process [34, 35] which is carried by weak and gravitational interactions. The current constraint on the the annihilation cross section of the DM is obtained by using the present DM abundance [36] which ensures a possibility to detect the DM. The other possibility is the experiments of the scattering of DM particles off atomic nuclei within a detector, the direct detection of DM, which has an upper limit of the order of $10^{-7} - 10^{-6}\, pb$ [37] for the WIMP-nucleon cross section.

The present work is devoted to the annihilation cross section of DM candidates in the Randall Sundrum RS1 background [38, 39] in which a graviscalar particle, the radion, arises and leads to the annihilation of the DM pair (see [40] for the minimal model). Furthermore, we switched on the radion -SM Higgs mixing coming from the curvature-scalar mixing and see that this mixing enhances the annihilation cross section up to the current observational results. This is the case that the SM Higgs also drives the annihilation in addition to the radion. For a second scenario, we consider that the unparticle is a DM candidate (see [41]) and both radion and Higgs fields drive the annihilation even there is no mixing between these two scalars. With the radion -SM Higgs mixing the physics becomes richer. Now we start to examine the annihilation cross section of the DM in these two scenarios:

# The DM as an additional scalar on the 3 brane

We consider the RS1 background and all particles, including the DM, live in the visible brane. In RS1 scenario, two 3 branes, the Planck brane and the TeV (visible) brane are the boundaries of the 5D world, which is compactified into $S^1/Z_2$ orbifold. This scenario ensures a solution to the well known hierarchy problem with the assumption that the gravity is concentrated near the Planck brane and extends into the bulk with varying strength. In addition to this the low energy effective theory has flat 4D space-time since vanishing 5D cosmological constant in both branes have equal and opposite tensions. The metric of the RS1 background reads

$$ds^2 = e^{-2\,A(y)}\,\eta_{\mu\nu}\,dx^\mu\,dx^\nu - dy^2\,, \tag{1}$$

where $A(y) = k\,|y|$, k is the bulk curvature constant, y is the extra dimension parametrized as $y = R\,\theta$ and the exponential factor $e^{-k\,L}$ with $L = R\,\pi$, is the warp factor which rescales the mass terms in order to the bring down the TeV scale, with a rough estimate of $L \sim 30/k$. Here, a scalar field, so called the radion $r$ field, is introduced as the fluctuation over the expectation value of the field $L(x)$, the size $L$. The equivalence principle leads to a mass to the field $L(x)$ and a stabilization mechanism for $r$ was proposed by Goldberger and Wise [42]. Finally, the metric in 5D is defined as [43]

$$ds^2 = e^{-2\,A(y)-2\,F(x)}\,\eta_{\mu\nu}\,dx^\mu\,dx^\nu - (1 + 2\,F(x))\,dy^2\,, \tag{2}$$

where the scalar field $F(x)$ reads,

$$F(x) = \frac{1}{\sqrt{6}\,M_{Pl}\,e^{-k\,L}}\,r(x)\,. \tag{3}$$

with normalized radion field, $r(x)$, (see [44]). Finally the induced metric at the orbifold point $\theta = \pi$ (visible brane) reads,

$$g^{ind}_{\mu\nu} = e^{-2\,A(L)-2\frac{\gamma}{v}\,r(x)}\,\eta_{\mu\nu}\,. \tag{4}$$

where $\gamma = \frac{v}{\Lambda_R}$, $\Lambda_R = \sqrt{6}\,M_{Pl}\,e^{-k\,L}$ and $v$ is the vacuum expectation value of the SM Higgs boson. Now, we introduce an additional scalar SM singlet field $\phi_S$, which was considered first by Silveria [45] and studied by several authors [46]-[50], and consider the action obeying the $Z_2$ symmetry $\phi_S \to -\phi_S$

$$\mathcal{S}_S = \int d^4x \sqrt{-g^{ind}} \left( \frac{1}{2}\,g^{ind\,\mu\nu}\partial_\mu\,\phi_S\,\partial_\nu\,\phi_S - \frac{1}{2}\,m_S^2\,\phi_S^2 \right)\,, \tag{5}$$

where $\sqrt{-g^{ind}} = e^{-4A(L) - 4\frac{\gamma}{v} r(x)}$ is the determinant of the induced metric on the visible brane. Notice that the term $e^{-4A(L)}$ in $\sqrt{-g^{ind}}$ is embedded into the redefinitions of the fields on the visible brane, namely, they are warped as $\phi_S \to e^{A(L)} \phi_{S\,warp}$, $m_S \to e^{A(L)} m_{S\,warp}$ and in the following we use warped fields without the *warp* index. The action eq.(5) leads to

$$\mathcal{S}'_S = \frac{1}{2} \int d^4 x \left( e^{-2\frac{\gamma}{v} r} \eta^{\mu\nu} \partial_\mu \phi_S \partial_\nu \phi_S - e^{-4\frac{\gamma}{v} r} m_S^2 \phi_S^2 \right), \tag{6}$$

which is responsible for the annihilation of the DM $\phi_S$, driven by the $\phi_S \phi_S r$ vertex,

$$V_1 = i \frac{\gamma}{v} (s + 2 m_S^2) \tag{7}$$

with $s = 4 m_S^2$. At this stage we consider that the curvature-scalar interaction

$$\mathcal{S}_\xi = \int d^4 x \sqrt{-g^{ind}} \, \xi \, \mathcal{R} \, H^\dagger H \tag{8}$$

is switched on where $H$ is the Higgs scalar field

$$H = \frac{1}{\sqrt{2}} (v + H^0), \quad <H> = \frac{v}{\sqrt{2}}, \tag{9}$$

and $\xi$ is the positive parameter. The interaction in eq.(8) results in the radion-SM Higgs mixing [51]-[56] and the mass eigenstates $H_p$ and $r_p$ become mediators of the DM annihilation process (see appendix A for brief review). Here the $\phi_S \phi_S H_p$ vertex arises after the mixing and the vertex factor of $\phi_S \phi_S H_p$ ($\phi_S \phi_S r_p$) interaction reads $b V_1$ ($a V_1$) where $V_1$ is given in eq.(7), $a$ ($b$) is the mixing parameter (see appendix A). Now, we present the total averaging annihilation rate of the DM, including the mixing effect:

$$<\sigma v_r> = \frac{4 |V_1|^2}{m_S} \left| \frac{b(d + \gamma b)}{s - m_{H_p}^2 + i m_{H_p} \Gamma_{H_p}} + \frac{a(c + \gamma a)}{s - m_{r_p}^2 + i m_{r_p} \Gamma_{r_p}} \right|^2 \Gamma'(\tilde{h} \to X_{SM}) + F_{\gamma g}, \tag{10}$$

where $c$ ($d$) is the mixing parameter (see appendix A), $\Gamma'(\tilde{h} \to X_{SM}) = \sum_{i=f,W,Z} \Gamma(\tilde{h} \to X_{i\,SM})$, with virtual Higgs $\tilde{h}$ having mass $2 m_S$ (see [57, 58]) and $v_r = \frac{2 p_{CM}}{m_S}$ is the average relative speed of two DM scalars (see for example [49]). The function $F_{\gamma g}$[1] is the contribution due to the $\gamma\gamma$ and $gg$ outputs and it reads

$$F_{\gamma g} = \frac{|V_1|^2 s^{\frac{3}{2}}}{16 \pi m_S} \left( \left| \frac{b \, c_\gamma^H}{s - m_{H_p}^2 + i m_{H_p} \Gamma_{H_p}} + \frac{a \, c_\gamma^r}{s - m_{r_p}^2 + i m_{r_p} \Gamma_{r_p}} \right|^2 \right.$$
$$\left. + 8 \left| \frac{b \, c_g^H}{s - m_{H_p}^2 + i m_{H_p} \Gamma_{H_p}} + \frac{a \, c_g^r}{s - m_{r_p}^2 + i m_{r_p} \Gamma_{r_p}} \right|^2 \right), \tag{11}$$

---

[1] For $ff$, $WW$, $ZZ$ outputs a common mixing factor appears for $H_p$ ($r_p$) mediation and the averaging annihilation rate can be written as a sum of corresponding decay widths, namely $\Gamma'(\tilde{h} \to X_{SM})$. However, for two photon ($\gamma\gamma$) and two gluon ($gg$) output the additional contribution coming from the trace anomaly (the terms $b_Y$, $b_2$ for $\gamma\gamma$ output and $b_{QCD}$ for $gg$ output) for the intermediate radion case (see for example [51, 54]) results in that one can not construct the part of the annihilation rate for $\gamma\gamma$ and $gg$ outputs in the form proportional to $\sum_{i=\gamma,g} \Gamma(\tilde{h} \to X_{i\,SM})$. Notice that $F_{\gamma g}$ can be written in the form of the first term in eq.(10) where $X_{SM} = \gamma, g$ when the terms coming from trace anomaly are ignored.

where the functions $c_\gamma^H$, $c_\gamma^r$, $c_g^H$ and $c_g^r$ are given in appendix B.

## Unparticle DM on the 3 brane

We consider unparticle as a DM candidate (see [41]) in the case that it obtains mass due to the interaction with the SM Higgs. Unparticles, which are the new degrees of freedom, has been proposed by Georgi [59, 60]. The starting point is a scale invariant hidden sector beyond the SM with non-trivial infrared fixed point. At low energy, around $\Lambda_U \sim 1\, TeV$, the hidden sector appears as unparticles which looks like a number of $d_U$ massless invisible particles where $d_U$ is the non-integer scaling dimension. In the low energy effective level the possible interactions between the SM particles and unparticles are described by the effective lagrangian (see for example [61]). Now we consider the action [41] having $Z_2$ symmetry for unparticle as

$$\mathcal{S}_U = -\int d^4x \sqrt{-g^{ind}} \frac{\lambda}{\Lambda_U^{2\,du-2}} U^2 H^\dagger H\,. \tag{12}$$

After the Higgs doublet develops the vacuum expectation value we get

$$\mathcal{S}'_U = -\frac{1}{2}\int d^4x \sqrt{-g^{ind}} \frac{\lambda}{\Lambda_U^{2\,du-2}} U^2 \left(H^{0\,2} + 2\,v\,H^0 + v^2\right), \tag{13}$$

where $\sqrt{-g^{ind}} = e^{-4\,A(L) - 4\frac{\gamma}{v} r(x)}$. By redefining the fields $\frac{U}{\Lambda_U^{du-1}}, H$ as $\frac{U}{\Lambda_U^{du-1}} \to e^{A(L)} (\frac{U}{\Lambda_U^{du-1}})_{warp}$, $H \to e^{A(L)} H_{warp}{}^2$, expanding $\sqrt{-g^{ind}} = e^{-4\frac{\gamma}{v} r(x)}$ and introducing the unparticle mass

$$m_U = \left(\frac{\sqrt{\lambda}\,v}{\Lambda_U^{du-1}}\right)^{\frac{1}{2-d_U}}, \tag{14}$$

we get the interaction term

$$\mathcal{L}'_U = -\frac{m_U^{4-2\,d_U}}{v} U^2 H^0 + \frac{2\,\gamma}{v} m_U^{4-2\,d_U} U^2\, r\,, \tag{15}$$

which is responsible for the annihilation of the DM, unparticle $U$ in this case[3]. Here the intermediate Higgs $H^0$ and radion $r$ are responsible for the annihilation of the DM driven by the $UUH^0$ and $UUr$ vertices:

$$V_h = -i\frac{m_U^{4-2\,d_U}}{v}\,,\ V_r = i\frac{2\,\gamma}{v} m_U^{4-2\,d_U}\,. \tag{16}$$

If we switch on the Higgs-radion mixing by considering the curvature-scalar interaction eq.(8), the DM annihilation is carried by the mass eigenstates $H_p$ and $r_p$ with the $UUH_p$ and $UUr_p$ vertices having the strengths $bV_r + dV_h$ and $aV_r + cV_h$ where $V_h$, $V_r$ are given in eq.(16) and $a$, $b$, $c$, $d$ are the mixing parameters (see appendix A). Finally the total averaging annihilation rate of the DM, including the mixing effect reads

$$<\sigma v_r> = \frac{4}{m_U} \left| \frac{(bV_r + dV_h)(d + \gamma b)}{s - m_{H_p}^2 + i m_{H_p} \Gamma_{H_p}} + \frac{(aV_r + cV_h)(c + \gamma a)}{s - m_{r_p}^2 + i m_{r_p} \Gamma_{r_p}} \right|^2 \Gamma'(\tilde{h} \to X_{SM}) + G_{\gamma g}, \quad (17)$$

where $\Gamma'(\tilde{h} \to X_{SM}) = \sum_{i=f,W,Z} \Gamma(\tilde{h} \to X_{i\,SM})$, virtual Higgs $\tilde{h}$ having mass $2 m_U$, $v_r = \frac{2 p_{CM}}{m_U}$ and $s = 4 m_U^2$. Similar to the previous case the function $G_{\gamma g}$ is the contribution due to $\gamma\gamma$ and $gg$ outputs and it reads

$$\begin{aligned} G_{\gamma g} &= \frac{s^{\frac{3}{2}}}{16\pi m_U} \left( \left| \frac{(bV_r + dV_h) c_\gamma^H}{s - m_{H_p}^2 + i m_{H_p} \Gamma_{H_p}} + \frac{(aV_r + cV_h) c_\gamma^r}{s - m_{r_p}^2 + i m_{r_p} \Gamma_{r_p}} \right|^2 \right. \\ &+ \left. 8 \left| \frac{(bV_r + dV_h) c_g^H}{s - m_{H_p}^2 + i m_{H_p} \Gamma_{H_p}} + \frac{(aV_r + cV_h) c_g^r}{s - m_{r_p}^2 + i m_{r_p} \Gamma_{r_p}} \right|^2 \right), \end{aligned} \quad (18)$$

where the functions $c_\gamma^H$, $c_\gamma^r$, $c_g^H$ and $c_g^r$ are given in appendix B.

---
[2]In the following we use warped fields without the *warp* index.
[3]Notice that the stability of unparticle is ensured with the considered $Z_2$ symmetry.

## Discussion

The present total annihilation rate is restricted by using the DM abundance which is determined by the WMAP collaboration [36] and, at two sigma level, it reads

$$\Omega h^2 = 0.111 \pm 0.018. \quad (19)$$

The expression connecting the annihilation cross section to the relic density is

$$\Omega h^2 = \frac{x_f \, 10^{-11} \, GeV^{-2}}{<\sigma v_r>}, \quad (20)$$

with $x_f \sim 25$ (see for example [2, 21, 49, 62, 63]) leads to the bounds

$$<\sigma v_r> = 0.8 \pm 0.1 \, pb,$$

of the order of $(1-2) \times 10^{-9} \, GeV^{-2}$. This is the case that s-wave annihilation is dominant (see [64] for details.).

In the present work we study the annihilation cross section of DM candidates in the framework of the RS1 scenario. First, we consider a DM candidate, living in the 4D brane, with the action given in eq.(5). In this case the DM annihilation process is induced by the radion, which has a natural trilinear coupling with the DM. Here we also consider the possible mixing of the radion and Higgs fields, arising with the inclusion of the curvature-scalar mixing (eq.(8)) and we extend the set of mediating particles which induce the annihilation, namely, the radion and the SM Higgs field. Here, the trilinear coupling of the Higgs field to the DM matter is regulated by the radion DM DM coupling and by the strength of the mixing. Second, we assume that the unparticle, which gets mass driven by the interaction (see eq.(13)) with the Higgs field, is a DM candidate. Similar to the previous case we analyze the annihilation cross section, which is induced by the radion and the Higgs field by switching on the mixing of the radion and Higgs field. Notice that, in this case, both radion and Higgs fields drive the annihilation even there is no mixing between these two scalars.

In both scenarios there exist numerous free parameters which should be restricted in the numerical calculations. In the first scenario, the Higgs mass $m_{H^0}$, the radion mass $m_r$, the mixing parameter $\xi$, the scale $\Lambda_R$ and the DM mass $m_S$ are the free parameters. After the mixing two mass eigenstates $m_+$ and $m_-$ which are functions of $m_{H^0}$, $m_r$, $\xi$ and $\gamma$ arise. First (second) we choose the heavy-light (light-heavy) one as the physical SM Higgs-radion mass, namely $m_+ = m_{H_p}$- $m_- = m_{r_p}$ ($m_+ = m_{r_p}$- $m_- = m_{H_p}$), fix the mass eigenstate $m_{H_p}$, $m_{H_p} = 120\, GeV$, and take different values of $m_{r_p}$. For the mixing parameter we respect the theoretical restriction, given in eq.(25) and LEP/LEP2 constraints (see [53]). For the scale $\Lambda_R$, we choose two different numerical values, $1.0\, TeV$ and $5.0\, TeV$ and we take the DM mass in the range $10\, GeV \leq m_S \leq 60\, GeV$. In the second scenario the scale $\Lambda_U$, the scaling dimension $d_U$ and the interaction parameter $\lambda$, which we take as $\lambda < 1.0$ not to loose the perturbative behavior, are the additional free parameters and we choose the appropriate values by fixing the DM mass (see eq.(14)). In both scenarios we restrict the parameters not to face with a possible perturbative unitarity violation (see for example [65] for a discussion of perturbative unitarity). In the calculations we take the SM Higgs vacuum expectation value $v = 246\, GeV$ and respect the upper and lower bounds of the current experimental value of the relic abundance, namely $0.7\, pb \leq <\sigma v_r> \leq 0.9\, pb$. In each figure we show these upper and lower bounds as a pair of parallel solid lines.

In Fig.1 (2) we plot the DM mass $m_S$ dependence of the annihilation cross section $<\sigma v_r>$ for $m_{H^0} = 120\, GeV$ and $\Lambda_R = 5.0\, TeV$. Here the solid-dashed (solid-dashed-short dashed) line represents $<\sigma v_r>$ for $m_R = 80-60\, GeV$, $\xi = 0.8-1.5$ ($m_R = 130-140-150\, GeV$, $\xi = 0.20-0.75-0.80$). We observe that the annihilation cross section lies within the current limits when the DM mass is in the vicinity of the resonant annihilation mass. For heavy radion one needs more fine tuning of the DM mass in order to obtain the current limit of the annihilation cross section. On the other hand the mixing results in a broader region for the restriction of the DM mass and, if the mixing is switched off, the annihilation cross section almost vanishes for the case of heavy radion, since the DM mass, which is lying in the region $m_S < 60\,GeV$, is far from the mass of the resonant annihilation induced by the radion[4]. However, in both cases, for the light and the heavy radion, the DM mass should be restricted strongly in order to reach the annihilation cross section, even the mixing is switched on.

Fig.3(4) is devoted to the DM mass $m_S$ dependence of the annihilation cross section $<\sigma v_r>$ for $m_{H^0} = 120\,GeV$ and $\Lambda_R = 1.0\,TeV$. Here the solid-dashed-short dashed-dotted (solid-dashed-short dashed) line represents $<\sigma v_r>$ for $m_R = 80 - 80 - 60 - 60\,GeV$, $\xi = 0.15 - 0.00 - 0.15 - 0.00$ ($m_R = 130 - 140 - 150\,GeV$, $\xi = 0.10 - 0.10 - 0.15$). Fig.3 shows that the mixing relaxes the restriction of the DM mass and the range for the restriction increases more than three times compared the one obtained without the mixing effect.

In Fig.5 (6) we present the mixing parameter $\xi$ dependence of the annihilation cross section $<\sigma v_r>$ for $m_{H^0} = 120\,GeV$ and $\Lambda_R = 1.0\,TeV$. Here the solid-dashed-short dashed-dotted (wide solid-narrow solid-wide dashed-narrow dashed-wide short dashed-narrow short dashed) line represents $<\sigma v_r>$ for $m_R = 80\,GeV, m_S = 50\,GeV$-$m_R = 80\,GeV, m_S = 35\,GeV$-$m_R = 60\,GeV, m_S = 50\,GeV$-$m_R = 60\,GeV, m_S = 35\,GeV$ ($m_R = 130\,GeV, m_S = 50\,GeV$-$m_R = 130\,GeV, m_S = 55\,GeV$- $m_R = 140\,GeV, m_S = 50\,GeV$-$m_R = 140\,GeV, m_S = 55\,GeV$-$m_R = 150\,GeV, m_S = 50\,GeV$-$m_R = 150\,GeV, m_S = 55\,GeV$). The figures show that the increase in the mixing results in that the current annihilation cross section can be reached. In addition to this the DM mass must not be far from the resonant annihilation mass(es). (See for example the narrow solid, narrow dashed, narrow short dashed lines in Fig.6 where the resonant annihilation mass is $m_S = 60\,GeV$ for the considered DM mass range.)

Now, we study the annihilation cross section with the assumption that the DM candidate is unparticle having mass which arises by switching on the interaction with the SM Higgs field.

Fig.7 (8) represents the mixing parameter $\xi$ dependence of the annihilation cross section $<\sigma v_r>$ for $m_{H^0} = 120\,GeV$, $m_U = 55\,GeV$ and $\Lambda_R = 1.0\,TeV$. Here the solid-dashed-short dashed-dotted line represents $<\sigma v_r>$ for $m_R = 60\,GeV, d_U = 1.0$-$m_R = 60\,GeV, d_U = 1.1$-$m_R = 80\,GeV, d_U = 1.0$-$m_R = 80\,GeV, d_U = 1.1$ ($m_R = 140\,GeV, d_U = 1.0$-$m_R = 140\,GeV$, $d_U = 1.1$-$m_R = 150\,GeV, d_U = 1.0$-$m_R = 150\,GeV, d_U = 1.1$)[5]. It is observed that the current

---
[4]Notice that the SM Higgs does not appear as a mediating scalar in this case.
[5]Here we choose $\lambda$ in order to get the DM mass as $m_U = 55\,GeV$, namely for $d_U = 1.0$ $\lambda = 0.05$ and for

annihilation cross section can be reached if the scale dimension is near to one for strong mixing when one choose the DM mass near to the numeraical values $m_U = 55\,GeV$. For completeness we present the scale parameter $d_U$ dependence of the annihilation cross section $<\sigma v_r>$ in Fig.9 (10) for different values of the mixing parameter by restricting the DM mass in the range $10\,GeV < m_U < 60\,GeV$ and taking $m_{H^0} = 120\,GeV$ and $\Lambda_R = 1.0\,TeV$. Here the solid-dashed-short dashed line represents $<\sigma v_r>$ for $m_R = 80\,GeV$, $\xi = -0.2$, $\lambda = 0.2 - 0.4 - 0.6$ ($m_R = 150\,GeV$, $\xi = -0.15$, $\lambda = 0.2 - 0.5 - 0.8$). These figures show that the annihilation cross section strongly depends on the variation of the scale dimension $d_U$ and the restriction of $d_U$ becomes stronger with the increasing values of the interaction strength $\lambda$.

Finally, in Fig.11 (12) we present the interaction parameter $\lambda$ dependence of the annihilation cross section $<\sigma v_r>$ for $m_{H^0} = 120\,GeV$, $\Lambda_R = 1.0\,TeV$ and $10\,GeV < m_U < 60\,GeV$. Here the solid-dashed-short dashed line represents $<\sigma v_r>$ for $m_R = 80\,GeV$, $\xi = -0.2$, $d_U = 1.1 - 1.2 - 1.3$ ($m_R = 150\,GeV$, $\xi = -0.15$, $d_U = 1.1 - 1.3 - 1.5$). We observe that the annihilation cross section is strongly sensitive to the parameter $\lambda$ and this sensitivity increases with the increasing values of $\lambda$.

At this stage we would like to present our results:

- First we consider that the DM annihilation process is switched on with the radion mediation and the SM Higgs mediation appears with the curvature-scalar mixing. We observe that the mixing process plays an essential role in obtaining the DM annihilation cross section in the current range. Notice that the DM mass must not be far from the resonant annihilation mass(es) for both light and heavy radion cases.

- Second we consider the unparticle as a DM candidate with the mass term arising with the interaction given in eq.(13). In this case the mediators for the annihilation are the radion and the Higgs scalars even there is no mixing. We observe that the current annihilation cross section can be obtained by fine tuning of the free parameters, $d_U$, $\lambda$ and $\xi$, existing in the model.

With the forthcoming experimental measurements and more accurate observations it would be possible to understand the nature of the DM and to construct a theoretical background.

---

$d_U = 1.1$ $\lambda = 0.09$

# Appendix

## A  The curvature scalar mixing

The action given in eq.(8) result in a mixing between the SM Higgs field and the radion as

$$
\begin{aligned}
H^0 &= d\, H_p + c\, r_p, \\
r &= b\, H_p + a\, r_p,
\end{aligned}
\qquad (21)
$$

where $H_p$ and $r_p$ are the mass eigenstates the SM Higgs field and the radion. The parameters $a, b, c, d$ read

$$
\begin{aligned}
a &= \frac{cos\theta}{Z}, \\
b &= -\frac{sin\theta}{Z}, \\
c &= sin\theta + \frac{6\,\xi\,\gamma}{Z} cos\theta, \\
d &= cos\theta - \frac{6\,\xi\,\gamma}{Z} sin\theta,
\end{aligned}
\qquad (22)
$$

with

$$
Z^2 = 1 + 6\,\xi\,\gamma^2\,(1 - 6\,\xi), \qquad (23)
$$

and the mixing angle $\theta$ is

$$
tan 2\theta = 12\,\gamma\,\xi\,Z \frac{m_{H^0}^2}{m_r^2 + m_{H^0}^2 \left(36\,\xi^2\gamma^2 - Z^2\right)}. \qquad (24)
$$

Here one must have $Z^2 > 0$ in order to get a positive definite kinetic energy terms of $H_p$ and $r_p$ and this restriction leads to a natural constraint for the parameter $\xi$ as

$$
\frac{1}{12}\left(1 - \sqrt{1 + \frac{4}{\gamma^2}}\right) \leq \xi \leq \frac{1}{12}\left(1 + \sqrt{1 + \frac{4}{\gamma^2}}\right). \qquad (25)
$$

Finally the mass squared eigenvalues read

$$
m_\pm = \frac{1}{2\,Z^2}\left(\beta\,m_{H_0}^2 + m_r^2 \pm \sqrt{(\beta\,m_{H_0}^2 + m_r^2)^2 - 4\,m_{H_0}^2\,m_r^2\,Z^2}\right), \qquad (26)
$$

where $\beta = 1 + 6\,\xi\,\gamma^2$ and $m_{+(-)}$ is the greater (smaller) of the set $m_{H_p}$, $m_{r_p}$.

# B  Some functions appearing in the text

The functions $c_\gamma^H$, $c_\gamma^r$, $c_g^H$ and $c_g^r$ in eqs.(11) and (18) read

$$
\begin{aligned}
c_\gamma^H &= \frac{\alpha_e}{2\pi v}\left((d+\gamma b)\left(\frac{8}{3}x_{tH}\left(1+(1-x_{tH})f_{tH}\right)-(2+3\,x_{WH}+3\,x_{WH}(2-x_{WH})f_{WH})\right)\right.\\
&\quad\left.+\,b\,(b_2+b_Y)\,\gamma\right),\\
c_\gamma^r &= \frac{\alpha_e}{2\pi v}\left((c+\gamma a)\left(\frac{8}{3}x_{tr}\left(1+(1-x_{tr})f_{tr}\right)-(2+3\,x_{Wr}+3\,x_{Wr}(2-x_{Wr})f_{Wr})\right)\right.\\
&\quad\left.+\,a\,(b_2+b_Y)\,\gamma\right),\\
c_g^H &= \frac{\alpha_S}{2\pi v}\left((d+\gamma b)\,x_{tH}\left(1+(1-x_{tH})f_{tH}\right)+b\,b_{QCD}\,\gamma\right),\\
c_g^r &= \frac{\alpha_S}{2\pi v}\left((c+\gamma a)\,x_{tr}\left(1+(1-x_{tr})f_{tr}\right)+a\,b_{QCD}\,\gamma\right),
\end{aligned}
\qquad (27)
$$

where

$$
f_{iH(r)} = -\int_0^1 \frac{\ln[1-\frac{4y(1-y)}{x_{iH(r)}}]}{2y}, \qquad (28)
$$

with $x_{iH(r)} = \frac{4m_i^2}{m_{H_p(r_p)}^2}$ and $b_2 = \frac{19}{6}$, $b_Y = -\frac{41}{6}$, $b_{QCD} = 11 - \frac{2}{3}N_f$. Notice that in the calculation of the annihilation cross section $x_{iH(r)}$ is taken as $x_{iH(r)} = \frac{m_i^2}{m_S^2}$ where $m_S$ is the DM mass.

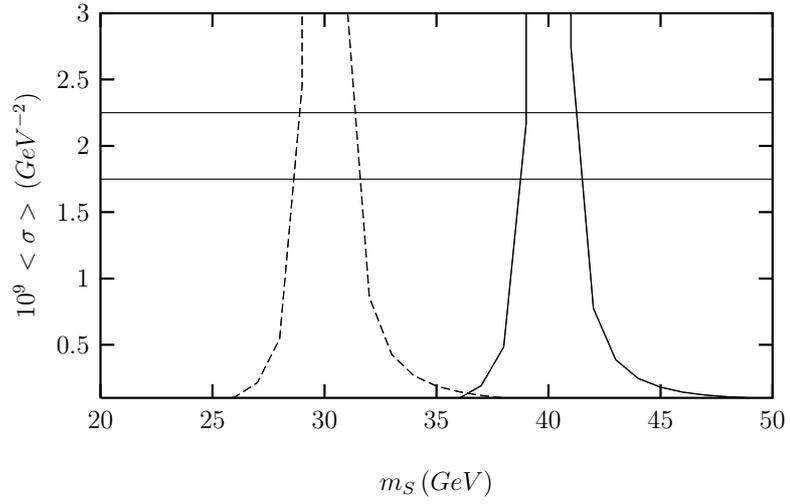

Figure 1: $<\sigma v_r>$ as a function of $m_S$ for $m_{H^0} = 120\,GeV$ and $\Lambda_R = 5.0\,TeV$. Here the solid-dashed line represents $<\sigma v_r>$ for $m_R = 80 - 60\,GeV$, $\xi = 0.8 - 1.5$

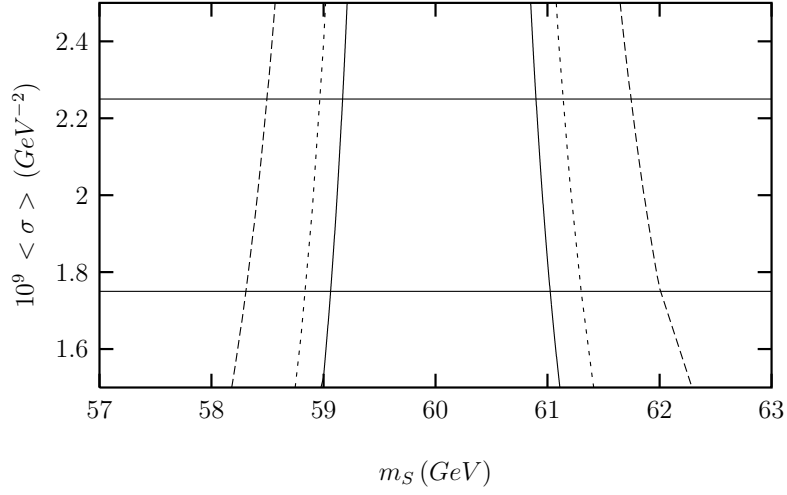

Figure 2: $<\sigma v_r>$ as a function of $m_S$ for $m_{H^0} = 120\,GeV$ and $\Lambda_R = 5.0\,TeV$. Here the solid-dashed-short dashed line represents $<\sigma v_r>$ for $m_R = 130 - 140 - 150\,GeV$, $\xi = 0.20 - 0.75 - 0.80$

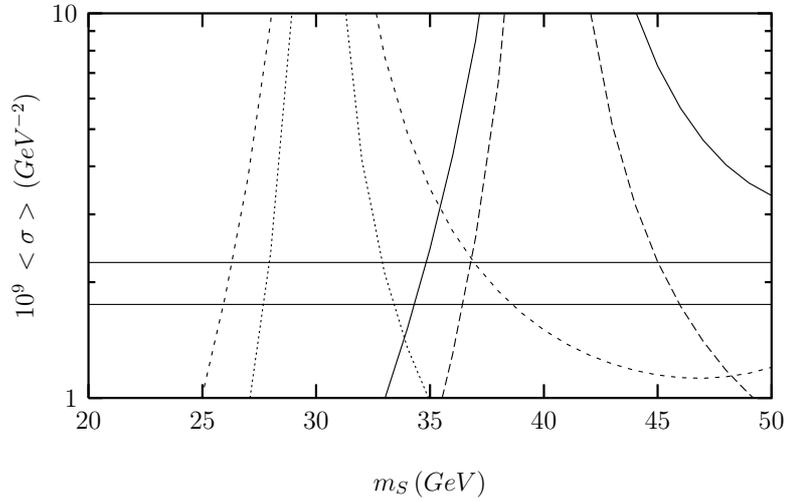

Figure 3: $<\sigma v_r>$ as a function of $m_S$ for $m_{H^0} = 120\,GeV$ and $\Lambda_R = 1.0\,TeV$. Here the solid-dashed-short dashed-dotted line represents $<\sigma v_r>$ for $m_R = 80 - 80 - 60 - 60\,GeV$, $\xi = 0.15 - 0.00 - 0.15 - 0.00$.

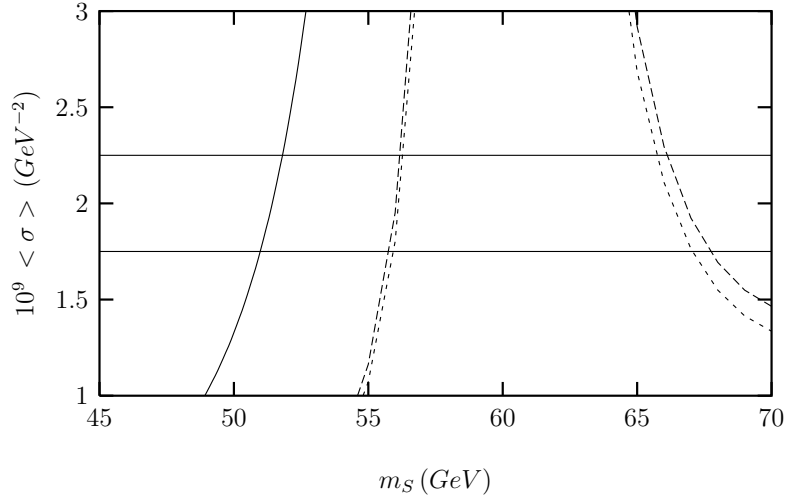

Figure 4: $<\sigma v_r>$ as a function of $m_S$ for $m_{H^0} = 120\,GeV$ and $\Lambda_R = 1.0\,TeV$. Here the solid-dashed-short dashed line represents $<\sigma v_r>$ for $m_R = 130 - 140 - 150\,GeV$, $\xi = 0.10 - 0.10 - 0.15$.

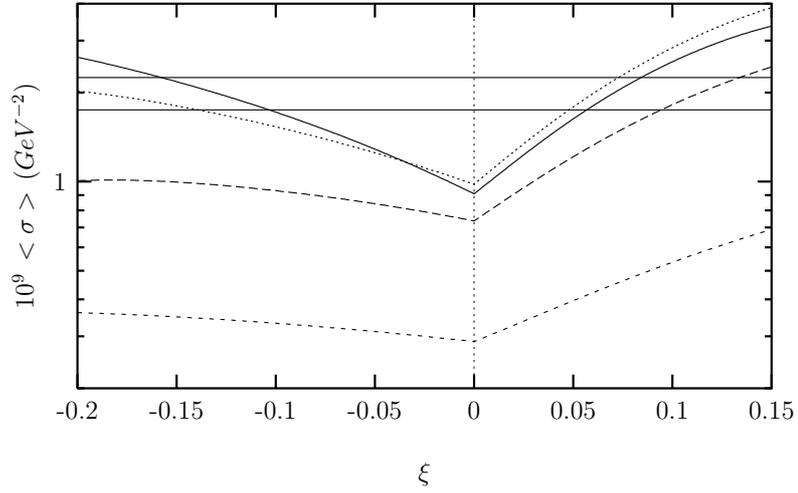

Figure 5: $<\sigma v_r>$ as a function of $\xi$ for $m_{H^0} = 120\,GeV$ and $\Lambda_R = 1.0\,TeV$. Here the solid-dashed-short dashed-dotted line represents $<\sigma v_r>$ for $m_R = 80\,GeV$, $m_S = 50\,GeV$- $m_R = 80\,GeV$, $m_S = 35\,GeV$- $m_R = 60\,GeV$, $m_S = 50\,GeV$-$m_R = 60\,GeV$, $m_S = 35\,GeV$.

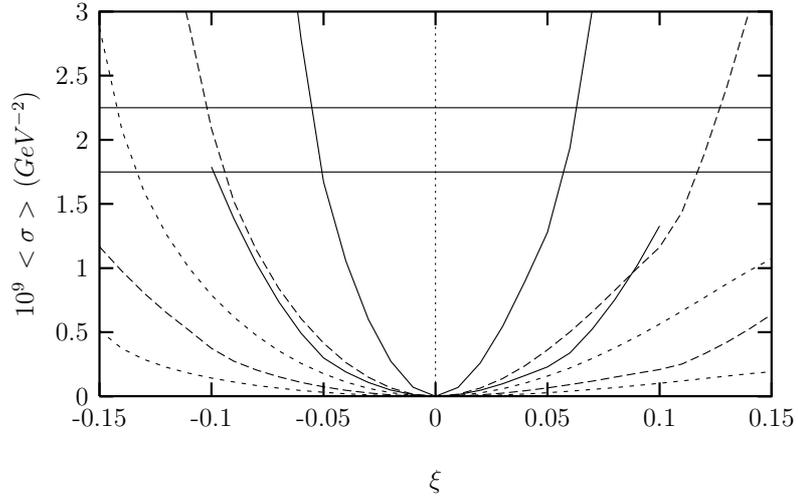

Figure 6: $<\sigma\,v_r>$ as a function of $\xi$ for $m_{H^0} = 120\,GeV$ and $\Lambda_R = 1.0\,TeV$. Here the wide solid-narrow solid-wide dashed-narrow dashed-wide short dashed-narrow short dashed line represents $<\sigma\,v_r>$ for $m_R = 130\,GeV$, $m_S = 50\,GeV$-$m_R = 130\,GeV$, $m_S = 55\,GeV$-$m_R = 140\,GeV$, $m_S = 50\,GeV$-$m_R = 140\,GeV$, $m_S = 55\,GeV$- $m_R = 150\,GeV$, $m_S = 50\,GeV$-$m_R = 150\,GeV$, $m_S = 55\,GeV$.

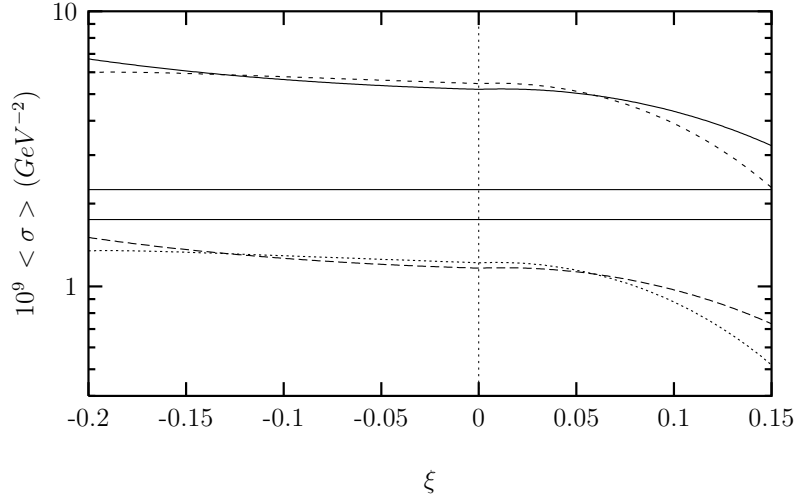

Figure 7: $<\sigma v_r>$ as a function of $\xi$ for $m_{H^0} = 120\,GeV$, $m_U = 55\,GeV$ and $\Lambda_R = 1.0\,TeV$. Here the solid-dashed-short dashed-dotted line represents $<\sigma v_r>$ for $m_R = 60\,GeV$, $d_U = 1.0$-$m_R = 60\,GeV$, $d_U = 1.1$-$m_R = 80\,GeV$, $d_U = 1.0$-$m_R = 80\,GeV$, $d_U = 1.1$.

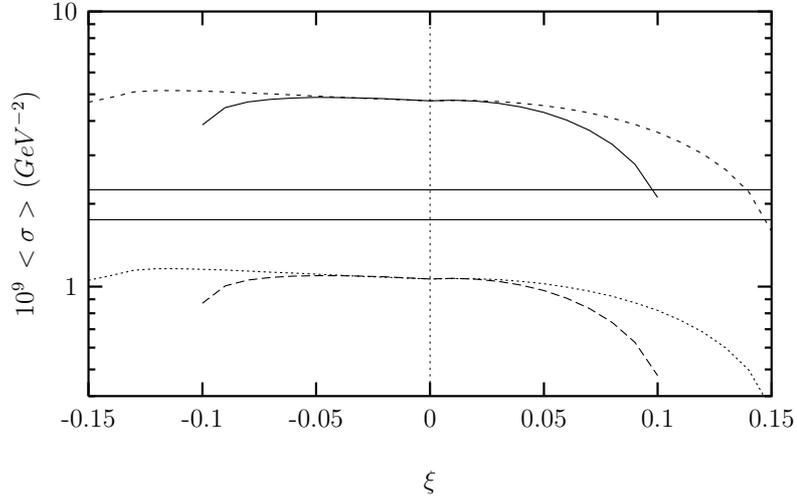

Figure 8: $<\sigma v_r>$ as a function of $\xi$ for $m_{H^0} = 120\,GeV$, $m_U = 55\,GeV$ and $\Lambda_R = 1.0\,TeV$. Here the solid-dashed-short dashed-dotted line represents $<\sigma v_r>$ for $m_R = 140\,GeV$, $d_U = 1.0$-$m_R = 140\,GeV$, $d_U = 1.1$-$m_R = 150\,GeV$, $d_U = 1.0$-$m_R = 150\,GeV$, $d_U = 1.1$.

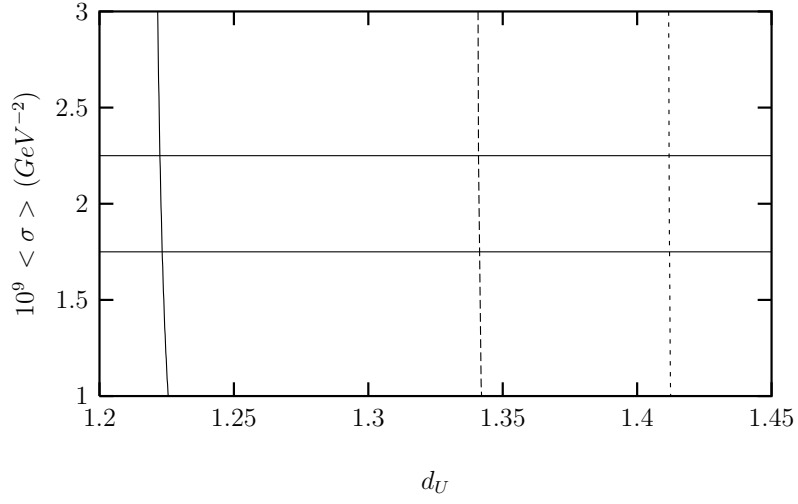

Figure 9: $<\sigma\, v_r>$ as a function of $d_U$ for $m_{H^0} = 120\, GeV$ and $\Lambda_R = 1.0\, TeV$. Here the solid-dashed-short dashed line represents $<\sigma\, v_r>$ for $m_R = 80\, GeV$, $\xi = -0.2$, $\lambda = 0.2 - 0.4 - 0.6$.

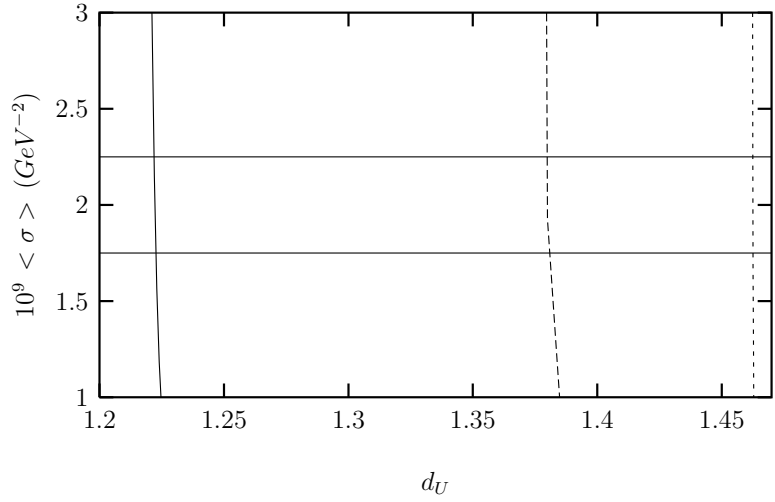

Figure 10: $<\sigma\, v_r>$ as a function of $d_U$ for $m_{H^0} = 120\, GeV$ and $\Lambda_R = 1.0\, TeV$. Here the solid-dashed-short dashed line represents $<\sigma\, v_r>$ for $m_R = 150\, GeV$, $\xi = -0.15$, $\lambda = 0.2 - 0.5 - 0.8$.

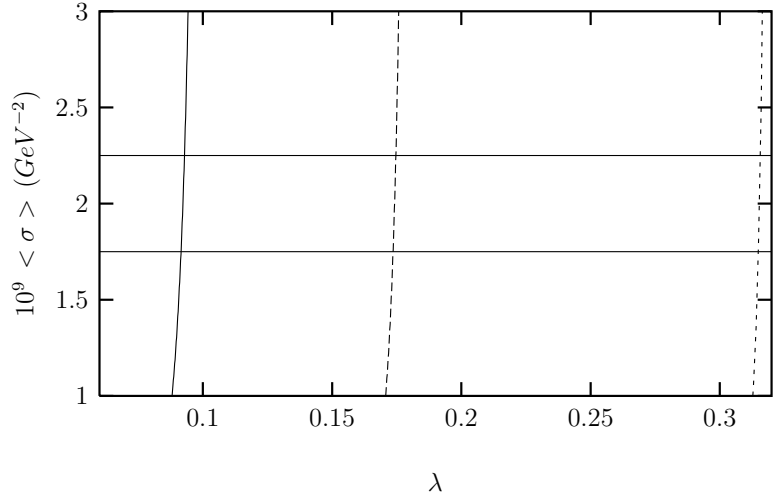

Figure 11: $<\sigma v_r>$ as a function of $\lambda$ for $m_{H^0} = 120\,GeV$, $\Lambda_R = 1.0\,TeV$ and $10 < m_U < 60\,GeV$. Here the solid-dashed-short dashed line represents $<\sigma v_r>$ for $m_R = 80\,GeV$, $\xi = -0.2$, $d_U = 1.1 - 1.2 - 1.3$.

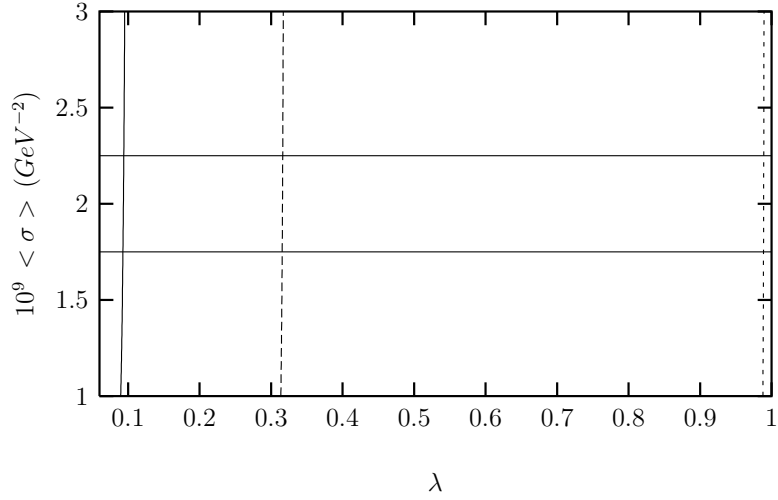

Figure 12: $<\sigma v_r>$ as a function of $\lambda$ for $m_{H^0} = 120\,GeV$, $\Lambda_R = 1.0\,TeV$ and $10 < m_U < 60\,GeV$. Here the solid-dashed-short dashed line represents $<\sigma v_r>$ for $m_R = 150\,GeV$, $\xi = -0.15$, $d_U = 1.1 - 1.3 - 1.5$.